\documentclass[a4paper,fleqn,usenatbib]{mnras}

\usepackage{amsmath}
\usepackage{amssymb}

\usepackage{mathptmx}
\usepackage{txfonts}

\usepackage[T1]{fontenc}
\usepackage{ae,aecompl}

\usepackage{graphicx}	

\title[Semi-analytic derivation of the threshold mass for prompt collapse in binary neutron star mergers]
{Semi-analytic derivation of the threshold mass for prompt collapse in binary neutron star mergers}

\author[Bauswein \& Stergioulas]{Andreas Bauswein$^{1}$, Nikolaos Stergioulas$^{2}$ \\
$^1$ Heidelberger Institut f\"ur Theoretische Studien, Schloss-Wolfsbrunnenweg 35, D-69118 Heidelberg, Germany \\
$^2$ Department of Physics, Aristotle University of Thessaloniki, GR-54124 Thessaloniki, Greece}

\pubyear{2017}

\begin{document}
\label{firstpage}
\pagerange{\pageref{firstpage}--\pageref{lastpage}}
\maketitle

\begin{abstract}
The threshold mass for prompt collapse in binary neutron star mergers was empirically found to depend on the stellar properties of the maximum-mass non-rotating neutron star model. Here we present a semi-analytic derivation of this empirical relation which suggests that it is rather insensitive to thermal effects, to deviations from axisymmetry and to the exact rotation law in merger remnants. We utilize axisymmetric, cold equilibrium models with differential rotation and determine the threshold mass for collapse from the comparison between an empirical relation that describes the angular momentum in the remnant for a given total binary mass and the sequence of rotating equilibrium models at the threshold to collapse (the latter assumed to be near the turning point of fixed-angular-momentum sequences). In spite of the various simplifying assumptions, the empirical relation for prompt collapse is reproduced with good accuracy, which demonstrates its robustness. We discuss implications of our methodology and results for understanding other empirical relations satisfied by neutron-star merger remnants that have been discovered by numerical simulations and that play a key role in constraining the high-density equation of state through gravitational-wave observations.
\end{abstract}



\section{Introduction}
Merging neutron stars (NSs) are the next type of source, which is expected to be detected with the current generation of gravitational-wave detectors. The outcome of a NS merger could be a black hole surrounded by an accretion torus (prompt collapse) or a massive rotating NS merger remnant. In the latter case the remnant may  undergo a gravitational collapse at a later time, as a result of  angular momentum redistribution and additional losses by mass ejection, neutrino emission and gravitational waves (see e.g.~\citet{2012LRR....15....8F} for a review).

The distinction between the prompt collapse scenario and the formation of a NS remnant is crucial for several observational aspects of NS mergers. These include the character and strength of the postmerger gravitational-wave emission, the amount of ejecta relevant for heavy element nucleosynthesis \citep{1977ApJ...213..225L,1989Natur.340..126E,1999ApJ...525L.121F} and nuclear powered electromagnetic emission \citep{1998ApJ...507L..59L,2005astro.ph.10256K,2010MNRAS.406.2650M}, and the conditions for the launch of a relativistic jet producing a short gamma-ray burst (GRB) \citep{1986ApJ...308L..43P,1989Natur.340..126E}.

The outcome of NS mergers (prompt collapse vs. NS remnant or delayed collapse) depends on the binary masses and the equation of state (EoS) of NS matter \citep[e.g.][]{2005PhRvL..94t1101S,2008PhRvD..78h4033B,2011PhRvD..83l4008H,2013PhRvL.111m1101B}. There are still significant uncertainties regarding the EoS of NS matter  and various theoretical models are available \citep[e.g.][]{2016PhR...621..127L,2016arXiv161003361O}. The prompt collapse to a black hole occurs for high total binary masses $M_\mathrm{tot}$, whereas less massive systems lead to the formation of an at least transiently stable merger remnant. For a given EoS, one can thus introduce a threshold binary mass $M_\mathrm{thres}$ that distinguishes the two different scenarios. For $M_\mathrm{tot}>M_\mathrm{thres}$  prompt collapse occurs, while $M_\mathrm{tot}<M_\mathrm{thres}$ results in a massive NS remnant that is stable for at least some number of dynamical timescales.

In previous work and within a systematic study of several representative EoSs we found that the threshold mass depends in a particular way on the EoS \citep{2013PhRvL.111m1101B}. The threshold binary mass can be described as a fraction $k$ of the maximum mass $M_\mathrm{max}$ of non-rotating NSs: $M_\mathrm{thres}=k\,M_\mathrm{max}$ with $k=k(C_{\rm max})$ scaling tightly with the maximum compactness $C_\mathrm{max}$ of non-rotating NSs. The maximum compactness is defined by $C_\mathrm{max}:=\frac{G M_\mathrm{max}}{c^2R_\mathrm{max}}$ with $R_\mathrm{max}$ being the radius of the maximum-mass configuration of non-rotating NSs, whereas  $G$ is the gravitational constant and  $c$ is the speed of light. Based on results from hydrodynamical simulations, $k(C_{\rm max}) $ can be fitted by $k=-3.38\,C_\mathrm{max}+2.43$ to good accuracy, which represents a purely empirical finding.

The unique relation between the threshold mass and properties of non-rotating NSs ($M_\mathrm{max}$ and $R_\mathrm{max}$) is important because it offers the opportunity to infer these quantities (which are directly related to the EoS) from observations. The threshold mass $M_\mathrm{thres}$ could be observationally constrained by measuring the total binary mass from the gravitational-wave inspiral signal of a NS merger and by testing for the presence or absence of postmerger gravitational-wave emission originating from a NS remnant (assuming that the detector would have the required sensitivity to detect a postmerger signal if there was one \citep{2014PhRvD..90f2004C}). Alternatively, if future theoretical models clarify the exact conditions leading to a short GRB, the observed electromagnetic emission may reveal whether or not in a given event a prompt collapse of the merger remnant occurred. In combination with a simultaneous gravitational-wave observation providing the binary masses, the threshold mass $M_\mathrm{thres}$ to black hole formation can be estimated. Similarly, observing a radioactively powered electromagnetic counterpart of a gravitational-wave detection may reveal the occurrence of a prompt collapse, since for equal-mass binaries direct black hole formation leads to smaller ejecta masses and thus different properties of the electromagnetic emission \citep{2013PhRvD..87b4001H,2013ApJ...773...78B}.

The radius $R_\mathrm{max}$ can be determined by measuring the dominant oscillation frequency of the postmerger phase for systems with binary masses slightly below $M_\mathrm{thres}$ \citep{2013PhRvL.111m1101B,2014PhRvD..90b3002B}. $R_\mathrm{max}$ can also be obtained by an extrapolation of the measured postmerger gravitational-wave frequencies of low-mass binary systems \citep[see][]{2014PhRvD..90b3002B,Bauswein:2015wsa,2016EPJA...52...56B}.

Given $M_\mathrm{thres}$ and $R_\mathrm{max}$, the maximum mass of non-rotating NSs could then be deduced by inverting the relation $M_\mathrm{thres}=k(C_\mathrm{max}) M_\mathrm{max}=M_\mathrm{thres}(M_\mathrm{max},R_\mathrm{max})$ describing the collapse behavior of merger remnants. We remark that the ratio $k$ can be similarly described as function of $M_\mathrm{max}/R_{1.6}$ with $R_{1.6}$ being the radius of a non-roating NS with a gravitational mass of 1.6~$M_\odot$ \citep{2013PhRvL.111m1101B}. Compared to $R_\mathrm{max}$ the radius $R_{1.6}$ may be easier to measure, e.g. by gravitational-wave detections \citep{2012PhRvD..86f3001B,2014PhRvD..90f2004C,2016CQGra..33h5003C}.  We also note that the relation between $k$ and $C_\mathrm{max}$ (Eq.~\ref{ksim}) has been found empirically through NS merger simulations for equal-mass binaries. For some candidate EoSs it has been verified that the same relation holds for slightly asymmetric binaries with mass ratios $q=M_1/M_2\approx 0.9$ with $M_1$ and $M_2$ being the masses of the binary compenents \citep{2013PhRvL.111m1101B}. A future measurement of the inspiral gravitational-wave signal of a NS merger will determine the mass ratio $q$ sufficiently accurate, \citep[e.g.][]{2014ApJ...784..119R,2016ApJ...825..116F}, to decide if the observed binary should follow the relation established for perfectly symmetric mergers. More general relations for arbitrary mass ratios still have to be determined through simulations although one may expect that the relation for symmetric binaries provides a fairly good estimate.

In this paper we provide a more general view on the stability of NS merger remnants by considering equilibrium models of rotating NSs. Using a simplified 
setup we corroborate in a more general context the specific dependence of the threshold mass on stellar parameters of non-rotating NSs. In doing this, we do not intend to construct equilibrium models that quantitatively resemble merger remnants to high accuracy. This would require significant fine-tuning and an extensive analysis of available hydrodynamical data. Instead, we aim at reproducing only the qualitative behavior with minimal assumptions. Such an approach
is important because it is independent of time-consuming and sophisticated hydrodynamical simulations,
whereas it may allow a first qualitative investigation of a large sample
of EoS models without employing computationally expensive calculations for
many different binary configurations.

Efforts to interpret equilibrium models of differentially rotating NSs in the context of merger remnants have been presented in e.g. \citet{2000ApJ...528L..29B,2003ApJ...583..410L,2004ApJ...610..941M,2012A&A...541A.156G,2014ApJ...790...19K,2016MNRAS.463.2667S,2016arXiv160902336G} (see \citet{2016arXiv161203050P}
for a review). Various studies have also considered rigidly rotating NSs as models for the late-time structure of merger remnants, when uniform rotation is enforced on a viscous or MRI timescale \citep[e.g.][]{2014PhRvD..89d7302L,2015ApJ...812...24F,2015ApJ...808..186L,2015ApJ...798...25D,2015PhRvL.115q1101M,2016PhRvD..93d4065G}.

The novelty of our approach lies in the fact that we relate equilibrium models to NS merger remnants by considering the detailed angular momentum budget provided by binary mergers as a function of mass. The relatively small computational demands of stellar equilibrium computations permit the investigation of a large number of different NS EoSs.

Finally, we note that the empirical relation that determines the available angular momentum in the merger remnant for a given total binary mass, should allow the construction of sequences of models resembling remnants of various masses. In turn, this will allow a detailed analysis of the oscillation modes of merger remnants which are relevant for the interpretation of postmerger gravitational-wave emission.

The paper is organized as follows. In Sect.~\ref{sec:setup} we describe the numerical method to compute equilibrium models and provide details on the employed EoSs as well as basic results from NS merger calculations. In the next section we discuss properties of differentially rotating stars and relate the results to the collapse behavior of NS mergers.

If not noted otherwise we use the term ``mass'' for the gravitational mass in isolation. If we refer to ``binary masses'', we mean the sum of the gravitational masses of the binary components at infinite binary separation. We work in geometrical units $G=c=1$ with the remaining scale set by $M_\odot=1$ if units are not explicitly mentioned.

\section{Setup and numerical method}\label{sec:setup}
\subsection{Stellar equilibrium code}
We use the RNS code \citep{1995ApJ...444..306S}  to construct axisymmetric equilibrium models of differentially rotating NSs \citep{2004MNRAS.352.1089S}, assuming a spacetime metric
of the form
\begin{equation}
  ds^2 = -e^{2 \nu} dt^2 + e^{2 \psi} (d \phi - \omega dt)^2 + e^{2 \mu}
  (dr^2+r^2 d \theta^2),
  \label{e:metric}
\end{equation}
where $\nu$, $\psi$, $\omega$ and $\mu $ are four metric functions
that depend on the coordinates $r$ and $\theta$ only. Matter is assumed to be a perfect fluid with stress energy tensor 
\begin{equation}
T^{\alpha\beta} = 
(e+P)u^\alpha u^\beta + P g^{\alpha\beta},
\end{equation}
where $\alpha, \beta$ are spacetime indices,  $g_{\alpha\beta}$ 
is the metric tensor, $u^\alpha$ is the four-velocity, $P$ is pressure and $e$ is energy density.  

We have extended the RNS code to a new, 3-parameter rotation law, that allows for a different rotational description of the envelope, compared to the core of the star.
Specifically, the usual 1-parameter rotation law introduced in \citet{1989MNRAS.237..355K} and used in a many previous studies (see \citet{2013rrs..book.....F,2016arXiv161203050P} for recent reviews) is extended as
\begin{equation}
F(\Omega) = \begin{cases}
               A_1^2(\Omega_c-\Omega)+(A_2^2-A_1^2)(1-\beta)\Omega_c, & \Omega\leq \beta\Omega_c, \\           
                A_2^2(\Omega_c-\Omega), &
               \beta\Omega_c \leq \Omega \leq \Omega_c,
            \end{cases}
            \label{rotlaw}
\end{equation}
where $F(\Omega):=u^tu_\phi$ and $A_1, A_2, \beta$ are the three parameters of the rotation law, while $\Omega:=u^t/u^\phi$ is the angular velocity as measured by an observer at infinity, $\Omega_c$ is the angular velocity at the center of the star.
The rotation law reduces to the usual 1-parameter law in  \citet{1989MNRAS.237..355K} when setting $\beta=1$. For the current qualitative study, we choose
$\hat A_1:=A_1/r_e=1,$  $\hat A_2:=A_2/r_e=1.225$ and $\beta=0.8$ with the equatorial radius $r_e$. 
Similar qualitative
behaviour is obtained for other values of the parameters that are within
the range that produces equilibrium models with similar bulk properties as
those of the remnants in simulations of binary NS mergers.

 We stress that since we are interested in the prompt collapse of remnants, we are not concerned with the detailed rotational profile several dynamical timescales after merging, which has been extracted e.g. in \citet{2005PhRvD..71h4021S,2015PhRvD..91f4027K,2016arXiv161008532G,2016arXiv161107152H,2016PhRvD..94d4060K,2016arXiv161203671K}. Hence, rotational law (\ref{rotlaw}) suffices for a first qualitative investigation such as the one presented here. In fact, our main result is rather insensitive to the details of the rotation law. It is only important to allow for a slower-rotating envelope such that stars can reach high masses (as those typical for remnants) without encountering mass-shedding. Further refinement of our findings can be performed in the future with more sophisticated rotation laws. 

\subsection{Equations of state}
\begin{figure}
   \includegraphics[width=8.8cm]{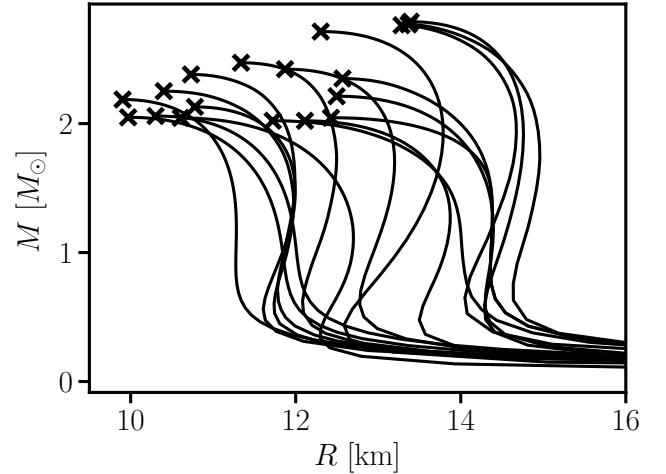} 
\caption{\label{fig:tov}Mass-radius relations of non-rotating NSs for all EoSs used in this study. $R$ denotes the circumferential radius and $M$ refers to the gravitational mass. Symbols mark the maximum-mass configurations.}
\end{figure}

For constructing equilibrium models of rotating
NSs, we are neglecting, to a first approximation, thermal effects, since we are
only interested in qualitatively reproducing the collapse behavior of merger
remnants. Remnants are in fact non-barotropic and constructing corresponding
equilibrium models would in any case require an averaging step, to produce
pseudo-barotropic equilibria \citep{2013rrs..book.....F}. However, for typical
temperatures of a few ten MeV as expected in merger remnants, the stellar
structure is only moderately altered at higher densities: at fixed density
the pressure is increased by order of 10 per cent compared to the pressure
at zero temperature (see e.g. Fig.~1 in \citet{2010PhRvD..82h4043B}). Hence,
the qualitative collapse behaviour is retained, to a first approximation,
even when considering zero-temperature EoSs\footnote{To asses the quantitative impact of thermal effects we redid hydrodynamical simulations for the DD2 EoS as in~\citet{2013PhRvL.111m1101B}. For this EoS we determined the threshold mass in simulations with the full temperature-dependent EoS table, in runs with the EoS at zero temperature and in calculations that employ an appximate treatment of thermal effects choosing different values of $\Gamma_\mathrm{th}$, which regulates the strength of the thermal pressure contribution. We find $M_\mathrm{thres}=3.35~M_\odot$ for the full table, $M_\mathrm{thres}=3.425~M_\odot$ for the zero-temperature calculation, and $M_\mathrm{thres}=3.35~M_\odot$ for $\Gamma_\mathrm{th}=1.5$, $M_\mathrm{thres}=3.425~M_\odot$ for $\Gamma_\mathrm{th}=1.75$ and $M_\mathrm{thres}=3.425~M_\odot$ for $\Gamma_\mathrm{th}=2$. We thus conclude that the influence of thermal effects on the collapse behavior is relatively small.}. Following the same arguments,
we assume neutrino-less beta-equilibrium to compute stellar equilibrium models.

We consider a wide range of a total of 18 EoSs. 8 of these EoSs
are available with full temperature and composition dependence
(DD2, LS220, LS375, NL3, SFHO, SFHX, TM1 and TMA, see
Table~\ref{tab1} for the definition of the acronyms and references), but are used in the zero-temperature limit for constructing equilibrium models. 8 EoSs (APR, ppAPR3, ppENG, ppH4, ppMPA1, ppMS1, ppMS1b and Sly4) are zero-temperature EoSs and are implemented in their piecewise polytropic form provided in \citet{2009PhRvD..79l4032R} (except for APR and SLy4, which are provided by tables taken from the Lorene package http://www.lorene.obspm.fr). Finally, we include two additional piecewise polytropes, where the parameters were chosen in order to obtain models with properties that are not covered by other EoSs. For these two additional piecewise polytropes  we set $\{\log{p_1}=34.75,\Gamma_1=3.0,\Gamma_2=2.0,\Gamma_3=2.0\}$ and $\{\log{p_1}=34.75,\Gamma_1=3.0,\Gamma_2=2.6,\Gamma_3=2.0\}$ in the terminology of \citet{2009PhRvD..79l4032R}. 

Table~\ref{tab1} lists the mass $M_\mathrm{max}$ and radius $R_\mathrm{max}$ of the maximum-mass configuration of non-rotating NSs described by these 18 EoSs (obtained by solving the Tolman-Oppenheimer-Volkoff (TOV) equations \citep{1939PhRv...55..364T,1939PhRv...55..374O}. All EoSs in our sample are compatible with the lower bound on the maximum mass $M_\mathrm{max}$ of non-rotating NSs set by the observation of NSs with a gravitational mass of $\sim 2~M_\odot$ \citep{2010Natur.467.1081D,2013Sci...340..448A}. The variety of NS properties within our sample of 18 EoSs is apparent from the mass-radius
relations of cold, non-rotating NSs displayed in Fig.~\ref{fig:tov}. The maximum mass $M_\mathrm{max}$ ranges between 2.02~$M_\odot$ and 2.79~$M_\odot$, while the radius of the maximum-mass configuration $R_\mathrm{max}$ ranges between 9.90~km and 13.39~km. The compactness $C_\mathrm{max}$ of the maximum-mass TOV configuration, which has been found to have a decisive impact on the collapse behavior of NS mergers, varies between 0.243 and 0.328 for the EoS models within our sample. 

For the first 8 EoSs which are available with full temperature and composition dependence
the threshold binary mass for prompt collapse to a black hole, $M_\mathrm{thres}^\mathrm{sim}$,
was determined by hydrodynamical
simulations in \citet{2013PhRvL.111m1101B} and is listed in Table~\ref{tab1}.
For the other EoSs in Table~\ref{tab1}, where $M_\mathrm{thres}^\mathrm{sim}$ is not listed, estimates for $M_\mathrm{thres}^\mathrm{sim}$ may be found in the literature, e.g. in \citet{2011PhRvD..83l4008H,2012PhRvD..86f3001B}. Note, however, that the approximate treatment of thermal effects, which is required for these zero-temperature EoSs, may lead to ambiguities in determining $M_\mathrm{thres}^\mathrm{sim}$.

\begin{table*}
	\centering
	\caption{EoSs with references used in this study. $M_\mathrm{max}$ and $R_\mathrm{max}$ refer to mass and radius of the maximum-mass configuration of non-rotating NSs. $M_\mathrm{thres}^\mathrm{eq}$ is the threshold mass to prompt collapse derived from equilibrium models (see text). $M_\mathrm{thres}^\mathrm{sim}$ denotes the threshold binary mass for prompt collapse determined by hydrodynamical simulations for fully temperature-dependent EoSs (data taken from~\citet{2013PhRvL.111m1101B}). For zero temperature EoSs the entry for $M_\mathrm{thres}^\mathrm{sim}$ is empty because these EoS models were not included in~\citet{2013PhRvL.111m1101B}. $R_\mathrm{thres}^\mathrm{eq}$ is the corresponding equatorial radius of the equilibrium model at the threshold to collapse with mass $M_\mathrm{thres}^\mathrm{eq}$.}
	\label{tab1}
	\begin{tabular}{|l|l|l|l|l|l|} 
	\hline
		EoS, references & $M_\mathrm{max}$ & $R_\mathrm{max}$ & $M_\mathrm{thres}^\mathrm{eq}$ & $M_\mathrm{thres}^\mathrm{sim}$ & $R_\mathrm{thres}^\mathrm{eq}$ \\         
         & $[M_\odot]$     & [km]        & $[M_\odot]$     & $[M_\odot]$  & [km]      \\  \hline
DD2,~\cite{2010PhRvC..81a5803T,2010NuPhA.837..210H}      & 2.42 &  11.87  & 3.24 & $\sim$3.35  & 15.91 \\  \hline
LS220,~\cite{1991NuPhA.535..331L}    & 2.04 &  10.61  & 2.94 & $\sim$3.05  & 14.53 \\  \hline
LS375,~\cite{1991NuPhA.535..331L}    & 2.71 &  12.30  & 3.39 & $\sim$3.65  & 16.53 \\  \hline
NL3,~\cite{1997PhRvC..55..540L,2010NuPhA.837..210H}      & 2.79 &  13.39  & 3.58 & $\sim$3.85  & 16.68 \\  \hline
SFHO,~\cite{2012arXiv1207.2184S}     & 2.06 &  10.30  & 2.86 & $\sim$2.95  & 14.08 \\  \hline
SFHX,~\cite{2012arXiv1207.2184S}     & 2.13 &  10.78  & 2.95 & $\sim$3.05  & 14.32 \\  \hline
TM1,~\cite{1994NuPhA.579..557S,2012ApJ...748...70H}      & 2.21 &  12.50  & 3.25 & $\sim$3.45  & 16.05 \\  \hline
TMA,~\cite{1995NuPhA.588..357T,2012ApJ...748...70H}      & 2.02 &  12.11  & 3.08 & $\sim$3.25  & 15.73 \\  \hline
APR,~\cite{1998PhRvC..58.1804A}      & 2.19 &  9.90   & 2.77 & -  & 13.92 \\  \hline
SLy4,~\cite{2001AA...380..151D}     & 2.05 &  9.97   & 2.81 & -  & 13.97 \\  \hline
ppAPR3,~\cite{1998PhRvC..58.1804A,2009PhRvD..79l4032R}   & 2.38 &  10.73  & 3.00 & -  & 14.60 \\  \hline
ppENG,~\cite{1996ApJ...469..794E,2009PhRvD..79l4032R}    & 2.25 &  10.40  & 2.93 & -  & 14.32 \\  \hline
ppH4,~\cite{2006PhRvD..73b4021L,2009PhRvD..79l4032R}     & 2.02 &  11.72  & 3.08 & -  & 15.48 \\  \hline
ppMPA1,~\cite{1987PhLB..199..469M,2009PhRvD..79l4032R}   & 2.47 &  11.34  & 3.15 & -  & 14.89 \\  \hline
ppMS1,~\cite{1996NuPhA.606..508M,2009PhRvD..79l4032R}    & 2.77 &  13.37  & 3.59 & -  & 16.85 \\  \hline
ppMS1b,~\cite{1996NuPhA.606..508M,2009PhRvD..79l4032R}   & 2.76 &  13.28  & 3.55 & -  & 16.95 \\  \hline
ppEoSa,~\cite{2009PhRvD..79l4032R}, this work   & 2.05 &  12.43  & 3.17 & -  & 16.40 \\  \hline 
ppEoSb,~\cite{2009PhRvD..79l4032R}, this work   & 2.35 &  12.57  & 3.32 & -  & 16.23 \\  \hline 

        \end{tabular}
\end{table*}

%
%
%
%
%
%
%
%

\subsection{Merger simulations}\label{ssec:merger}
As an input for selecting particular equilibrium models, we consider a number of merger simulations to construct an empirical relation between the angular momentum  of the remnant and the total binary mass.  We employ the simulation data from \citet{2014PhRvD..90b3002B}, where more details can be found. Since we are interested in the stability of the remnant directly after merging, we extract the angular momentum at the time of merging from simulations with different binary masses. For a given EoS the angular momentum $J_\mathrm{merger}$ depends linearly on the total binary mass $M_\mathrm{tot}$. For example, for the moderately stiff DD2 EoS this relation can be well described by the linear fit
\begin{equation}\label{eq:jmerger}
J_\mathrm{merger} \simeq a \,M_\mathrm{tot} - b,
\end{equation}
with $a=4.041$ and $b=4.658$ (see dashed line and squares in Fig.~\ref{fig:stab}). Below we will employ this fit to estimate the  angular momentum of the remnant for a given total binary mass. The fit~\eqref{eq:jmerger} is compatible with available data from other simulations (see, e.g., \citet{2016PhRvD..94b4023B}). We stress that, in what follows,  Eq.~\eqref{eq:jmerger} is the only input  we take from hydrodynamical merger simulations.

The NS EoS has a slight impact on the angular momentum of the merger remnant for a fixed binary mass. Stiffer EoSs lead to larger radii and to an increase of $J_\mathrm{merger}$ because the stars merge earlier compared to more compact NSs. Thus, less angular momentum is lost during the inspiral phase. Softer EoSs which yield more compact NSs result in a reduction of $J_\mathrm{merger}$. In essence, the relation~\eqref{eq:jmerger} is slightly shifted upwards or downwards depending on the EoS. The variation due to the EoS dependence in the angular momentum for a fixed total binary mass amounts to a few per cent. Hence, one can consider Eq.~\eqref{eq:jmerger} to be a practically EoS-independent empirical relation with a very small uncertainty.  

Still, if one wishes, the EoS dependence of the fit Eq.~\eqref{eq:jmerger} can be taken into account by a constant offset in the linear relation. To this end we consider the radius $R_{1.5}$ of a non-rotating NS with a gravitational mass of 1.5~$M_{\odot}$, which is a good measure for the compactness of the inspiralling NSs in mergers with total binary masses in the range between 2.4~$M_\odot$ and 3.6~$M_\odot$. We compare $R_{1.5}$ of a given EoS to $R_{1.5}^\mathrm{DD2}$, which is the corresponding radius of our reference model with the DD2 EoS (Eq.~\eqref{eq:jmerger}). Then we shift the fit parameter $b$ depending on the difference $R_{1.5}-R_{1.5}^\mathrm{DD2}$:
\begin{equation}
b\rightarrow b+\frac{R_{1.5}-R_{1.5}^\mathrm{DD2}}{2~km}*0.2.
\end{equation}
This choice is guided by a rough comparison to $J_\mathrm{merger}(M_\mathrm{tot})$ for other EoSs. We stress that the EoS effect on $J_\mathrm{merger}(M_\mathrm{tot})$ is at the level of a few per cent and has only a very small impact on our final results. Thus, the quantitative details of this correction are of minor importance, and we only incorporate this coarse procedure to verify the insensitivity of our final results to EoS variations in~\eqref{eq:jmerger}. 

\section{Results}\label{sec:results}
\subsection{Differentially rotating equilibrium models}
\begin{figure}
   \includegraphics[width=8.8cm]{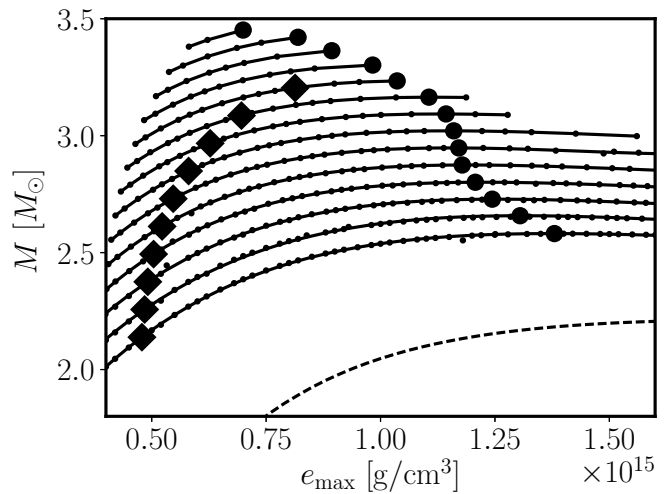}
\caption{\label{fig:jseq}Gravitational mass $M$ as function of the maximum energy density $e_{\rm max}$ of differentially rotating NSs with various fixed values of angular momentum $J$ for the TM1 EoS. The solid lines show fifth-order polynomial least-square fits to the sequences with $J={4,4.5,5,5.5,..., 10.5}$ (in geometrical units). Up to  $J=8.5,$ filled circles mark the turning point of each $J$-constant sequence. For the five highest values of $J$ the numerical code produces equilibrium sequences that only come close to the turning point (see text). In these cases the filled circles still represent a good approximation of the maximum mass that can be reached for each value of $J$ because the  $J$-constant sequences have a small slope. The filled diamonds are a sequence of models that satisfy the empirical relation \eqref{eq:jmerger} for binary NS merger remnants. The dashed line corresponds to the non-rotating limit.}
\end{figure}
Using the RNS code we compute differentially rotating equilibrium models for every EoS in Table I, varying systematically the central energy density $e_{\rm c}$ and polar to equatorial axis ratio, which are the parameters to be specified for obtaining a  stellar configuration. For every model we obtain the gravitational mass, the baryon mass, the angular momentum and other properties, such as the maximum energy density $e_{\rm max}$ (notice that in quasi-toroidal models $e_{\rm max}>e_c$, see \citet{2004MNRAS.352.1089S}). Based on a large number of computed equilibrium models we identify sequences of constant angular momentum $J$ for every EoS (employing linear interpolation). An example is shown in Fig.~\ref{fig:jseq} for the TM1 EoS. We display the gravitational mass as function of the maximum energy density for stars with fixed angular momenta of 4, 4.5, 5, 5.5, 6, 6.5, 7, 7.5, 8, 8.5, 9, 9.5, 10 and 10.5 (in geometrical units). Solid lines are fifth-order polynomial least square fits to the original data describing the sequences of constant angular momenta. Sequences with higher $J$ correspond to higher masses. For comparison, the dashed line shows the gravitational mass of the TOV solutions, i.e. the sequence of zero angular momentum.

For a given sequence of fixed $J$ the gravitational mass $M$ first increases with increasing maximum density and then decreases. In Fig.~\ref{fig:jseq} the filled circles mark the turning points of the $J$-constant sequences and we denote the maximum mass on a $J$-constant sequence as $M_\mathrm{stab}$. For uniformly rotating stars, a turning point along a  $J$-constant sequence marks the onset of the secular axisymmetric instability to collapse \citep{1988ApJ...325..722F} (models on the declining branch with $\frac{d M}{d e_\mathrm{max}}|_{J=\mathrm{const.}}<0$ are unstable), while the dynamical instability sets in nearby, see discussion in \citet{2013rrs..book.....F}.
For differentially rotating stars, there is no corresponding proof of a turning point theorem. However, the numerical findings in \citet{2014ApJ...790...19K}  imply that the location of the
dynamical instability to collapse, even in the case of differentially rotating models, may be  relatively near (within roughly 25\% in maximum density)\ to the turning points of
the $J$-constant curves
in Fig.~\ref{fig:jseq}. Since the $J$-constant curves have a small
slope over a relatively wide range of maximum densities for our choice of
moderate differential rotation, the  $J(M_\mathrm{stab})$ relation is rather insensitive to the precise value of the maximum density where the instability occurs. Thus, for our purposes, the turning points of
the $J$-constant curves
in Fig.~\ref{fig:jseq} serve as a way to estimate   $J(M_\mathrm{stab}$).

For high angular momenta (small axis ratio),  differentially rotating models can become quasi-toroidal or multiple types of differentially rotating models can exist for the same specified central density and axis ratio, see \citet{2004MNRAS.352.1089S,2007PhRvD..76b4019Z,2016MNRAS.463.2667S,2016arXiv160902336G}. The RNS code is not designed to distinguish between different types of models and is not converging in such cases (or when the axis ratio becomes smaller than $\sim0.15-0.25$, depending on the EoS, for quasi-toroidal models). Thus, for high angular momenta, the $J$-constant  sequences only reach {\it near} the turning point. In Fig.~\ref{fig:jseq} this occurs  for the sequences with $J=\{8.5,9,9.5,10,10.5 \}$, where the filled circles mark the configuration with the maximum mass along the obtained sequence. However, given that $\frac{d M}{d \rho}|_{J=\mathrm{const.}} \simeq 0$ near the turning point, the displayed filled circles still represent good approximations for the maximum masses of the respective sequences. In the following, we will thus use these approximate maximum masses as estimates of the masses $M_\mathrm{stab}$ at the stability limit, when an actual turning point is not determined. 

\subsection{Application to neutron-star mergers}
\subsubsection{Threshold mass to collapse}
\begin{figure}
  \includegraphics[width=8.8cm]{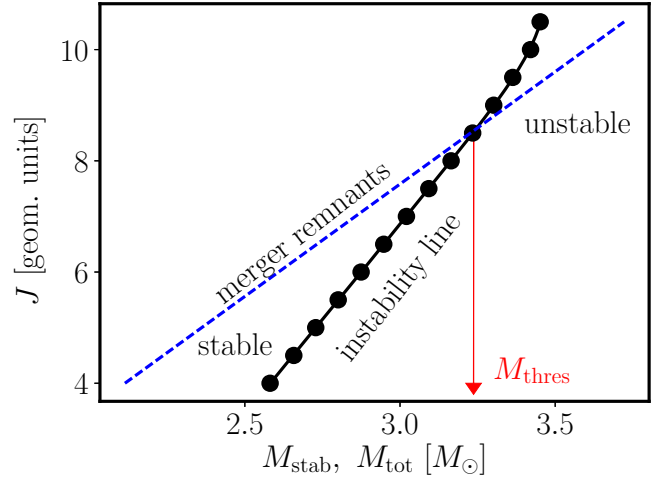}
\caption{\label{fig:stablimittm1}Angular momentum  $J(M_\mathrm{stab})$
of differentially rotating NSs as function of the maximum stable mass $M_\mathrm{stab}$  for a given amount of angular momentum computed for the TM1 EoS (filled circles and solid line). The filled circles correspond to the sequence of turning points (axisymmetric instability line) in Fig~\ref{fig:jseq}, while the solid line is a fifth-order polynomial least-squares fit to the data. The dashed line represents the angular momentum in merger remnants $J_\mathrm{merger}(M_\mathrm{tot}) $  as function of the total binary mass $M_\mathrm{tot}$. When $J_\mathrm{merger}(M_\mathrm{tot}) > J(M_\mathrm{stab})$ merger remnants are stable, otherwise they undergo prompt collapse to a black hole. The intersection $J_\mathrm{merger}(M_\mathrm{tot}) =J(M_\mathrm{stab}) $ defines the threshold mass $M_\mathrm{thres}$  for prompt collapse.}
\end{figure}

The importance of the turning points lies in the fact that they determine the maximum mass $M_\mathrm{stab}$ which can be supported against the gravitational collapse for a given amount of angular momentum. We display the relation $J(M_\mathrm{stab})$ between the angular momentum and the corresponding maximum stable mass in Fig.~\ref{fig:stablimittm1} for the TM1 EoS (the filled circles correspond to the same data as the filled circles in Fig.~\ref{fig:jseq}). The solid line represents a fifth-order least-square polynomial fit of these data. In addition, we display the empirical
relation for the angular momentum of binary NS merger remnants $J_\mathrm{merger}(M_\mathrm{tot}) $ as function of the total binary mass for the TM1 EoS (shown as a sequence of  filled diamonds  in Fig.~\ref{fig:jseq}  and as a dashed line in Fig.~\ref{fig:stablimittm1}; note that the location of the diamonds in Fig.~\ref{fig:jseq} are determined by the masses and angular momenta of the merger remnant and that the corresponding $e_\mathrm{max}$ do not have a direct physical interpretation). 

\begin{figure}
  \includegraphics[width=8.8cm]{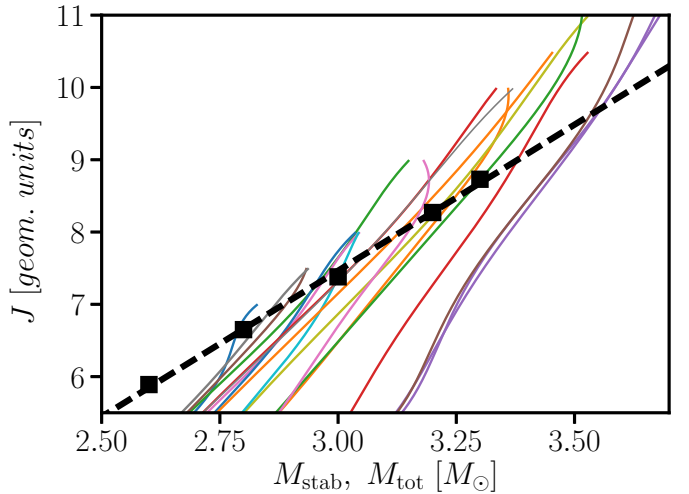}
\caption{\label{fig:stab}Angular momentum  $J(M_\mathrm{stab})$
of differentially rotating NSs as function of the maximum stable mass $M_\mathrm{stab}$
 for a given amount of angular momentum for different EoSs (solid lines of different colors). Figure similar to Fig.~\ref{fig:stablimittm1}. The filled squares are data points from merger simulations, representing the angular momentum in merger remnants $J_\mathrm{merger}(M_\mathrm{tot})
$  as function of the total binary mass $M_\mathrm{tot}$ (shown here using the DD2 EoS, but practically insensitive to the EoS), while the dashed line is a linear least-squares fit (Eq.~\eqref{eq:jmerger}). The intersections between the solid lines and the dashed line mark the threshold mass $M_\mathrm{thres}^\mathrm{eq}$ for prompt collapse for each EoS as explained in Fig.~\ref{fig:stablimittm1}.}
\end{figure}

\begin{figure}
  \includegraphics[width=8.8cm]{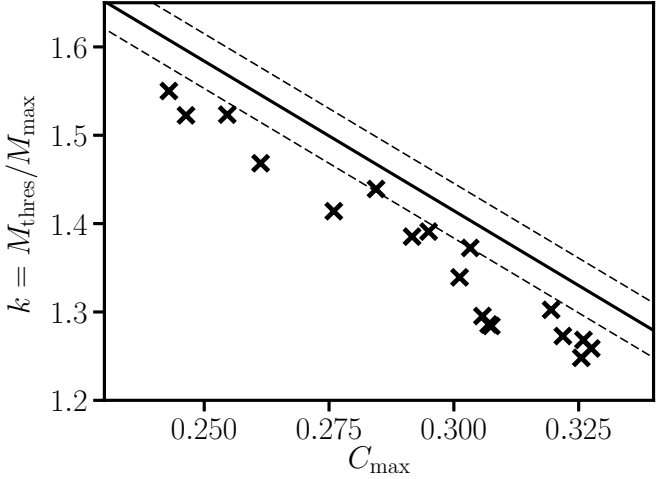}
\caption{\label{fig:kplot}The $\times$ symbols represent the ratio $k^\mathrm{eq}=M_\mathrm{thres}^\mathrm{eq}/M_\mathrm{max}$ as function of the compactness $C_\mathrm{max}$ of the maximum-mass configuration of non-rotating NSs for different EoSs. For comparison, the solid line represents a linear fit to the corresponding ratio $k^\mathrm{sim}=M_\mathrm{thres}^\mathrm{sim}/M_\mathrm{max}$ found through dynamical merger simulations. The dashed lines illustrate maximum error bars of the linear fit for $k^\mathrm{sim}$. Compare to Fig.~1 in \citet{2013PhRvL.111m1101B}.}
\end{figure}

The intersection between the two lines in Fig.~\ref{fig:stablimittm1}.
can now serve as a qualitative description of the threshold mass to collapse
in binary NS mergers. A NS merger with a small total binary mass leads to a merger remnant with an angular momentum exceeding what is necessary to support an equilibrium stellar model of the same mass (the dashed curve is above the solid curve in Fig.~\ref{fig:stablimittm1}). Within our approximations, the merger remnant would then be stable (at least on a secular timescale). On the other hand, for mergers with high total binary mass, the angular momentum inherited by the merger remnant is smaller than the angular momentum required to stabilize an equally massive equilibrium model (the dashed curve is below the solid curve in Fig.~\ref{fig:stablimittm1}). Within our approximations,
this would lead to a prompt collapse of the merger remnant to a black hole. 

The intersection between the solid curve and the dashed curve in Fig.~\ref{fig:stablimittm1} thus defines a ``theoretical'' threshold mass to prompt collapse. For the example with the TM1 EoS in this figure, the theoretically expected threshold mass is 3.24~$M_\odot$\footnote{Note that this value of the theoretically expected threshold mass slightly differs from the value listed in Tab.~\ref{tab1} because in Fig.~\ref{fig:stablimittm1} we employ data from hydrodynamical simulations for the TM1 EoS to describe $J_\mathrm{merger}(M_\mathrm{tot})$ instead of using the fit to the results with the DD2 EoS and the EoS-dependent correction procedure for $J_\mathrm{merger}$ as described in Sect.~\ref{ssec:merger}.}. For comparison, the actual threshold binary mass which has been found in our hydrodynamical simulations is  $\sim3.45~M_\odot$ for this EoS (see~\citet{2013PhRvL.111m1101B} and Table~\ref{tab1}). The two values differ by $\sim 6\%$. Given the various assumptions that we made when constructing the equilibrium models (axisymmetry, stationarity, a simple differential rotation law, zero-temperature EoS)
our qualitative derivation of the threshold mass agrees rather well with the actual numerical value.

Following the procedure described above we compute a ``theoretical'' threshold mass for every EoS in our sample (Table~\ref{tab1}) by determining the intersection between  $J(M_\mathrm{stab})$ for each EoS with the empirical relation $J_\mathrm{merger}(M_\mathrm{tot})$. The results are summarized in Fig.~\ref{fig:stab}. Solid curves show the fifth order polynomial fits describing  $J(M_\mathrm{stab})$, which we obtain from equilibrium models for different EoSs. The dashed curve displays the empirical relation $J_\mathrm{merger}(M_\mathrm{tot})$, with parameters obtained for the DD2 EoS, as in Eq.~\eqref{eq:jmerger}. In practice,  we also apply the small EoS-dependent correction to $J_\mathrm{merger}(M_\mathrm{tot})$ described in Section~\ref{ssec:merger}, when extracting
the intersection point between  $J(M_\mathrm{stab})$ and  $J_\mathrm{merger}(M_\mathrm{tot})$
for each EoS.
We denote the threshold mass as determined from equilibrium models as $M_\mathrm{thres}^\mathrm{eq}$, in contrast to $M_\mathrm{thres}^\mathrm{sim}$ extracted directly from merger simulations.

Table~\ref{tab1} lists  $M_\mathrm{thres}^\mathrm{eq}$ and $M_\mathrm{thres}^\mathrm{sim}$ (when available) for the 18 different EoSs we consider in the present work.
In all cases, our qualitative determination of  $M_\mathrm{thres}^\mathrm{eq}$ systematically underestimates $M_\mathrm{thres}^\mathrm{sim}$ by a relative difference of only $\sim 3-7\%$, which shows that our approximations still allow for a useful estimate of the threshold mass using only equilibrium models and avoiding time consuming simulations. 

Given that the  $J_\mathrm{merger}(M_\mathrm{tot})$ relation is very weakly
depending on the EoS, we conclude that the EoS dependence of the  threshold
mass $M_\mathrm{thres}$ is dominated by the EoS dependence of $J(M_\mathrm{stab})$.
(Not applying the EoS correction to $J_\mathrm{merger}(M_\mathrm{tot})$ leads to deviations of less than 2 per cent in $M_\mathrm{thres}^\mathrm{eq}$.)

\subsubsection{The $k=k(C_{\rm max})$ relation}
For NS mergers one can introduce the ratio $k^\mathrm{sim}:=M_\mathrm{thres}^\mathrm{sim}/M_\mathrm{max}$
of the binary threshold mass for direct black-hole formation to the
maximum mass of non-rotating NSs. The ratio $k^\mathrm{sim}$ has been found to be approximately a linear function
of $C_\mathrm{max}:=\frac{GM_\mathrm{max}}{c^2R_\mathrm{max}}$ \citep{2013PhRvL.111m1101B,2016EPJA...52...56B}. The simulation data can be well described by the linear relation 
\begin{equation}
k^{\rm sim}:=M_\mathrm{thres}^\mathrm{sim}/M_{\rm max}= 
-3.38\, C_\mathrm{max} + 2.43,
\label{ksim}
\end{equation} 
which is shown as solid line in Fig.~\ref{fig:kplot}. The thin dashed lines illustrate the maximum deviation of the simulation data from
the fit: all models considered in \citet{2013PhRvL.111m1101B} lie within the
band defined by the dashed lines.

In Fig.~\ref{fig:kplot} we also display the corresponding ratio $k^\mathrm{eq}:=M_\mathrm{thres}^\mathrm{eq}/M_\mathrm{max}$ as a function of $C_\mathrm{max}$ using now the threshold mass determined by the equilibrium models ($\times$ symbols) for various EoSs. Remarkably, using our equilibrium models we can reproduce the empirical relation (\ref{ksim}) the with very good accuracy. The data points in Fig.~\ref{fig:kplot}  that were produced using the equilibrium models can be described by the linear fit 
\begin{equation}
k^{\rm eq}:=M_\mathrm{thres}^\mathrm{eq}/M_{\rm max}= 
-3.49\, C_\mathrm{max} + 2.40,\label{eq:keq}
\end{equation} 
which has a similar maximum deviation as (\ref{ksim}). The slope differs by $\sim3\%$  with respect to the slope in (\ref{ksim}) and the constant term differs by only $1.4\%$.  The various approximations we made in our method using the equilibrium models (instead of time consuming simulations) thus only have small effects on the predicted relation   $k(C_\mathrm{max})$. Further refinement of our method (a different choice of rotation profile and inclusion of thermal effects) may yield an even more accurate agreement. 

We note that omitting the small correction for the EoS dependence of $J_\mathrm{merger}(M_\mathrm{tot})$ described in Section~\ref{ssec:merger}, and using Eq.~\eqref{eq:jmerger} with parameters obtained
for the DD2 EoS only, affects the ratio  $k^\mathrm{eq}$ by $<2\%$. 

We conclude that the existence of the empirical, linear $k(C_\mathrm{max})$ relation discovered numerically in \citet{2013PhRvL.111m1101B,2016EPJA...52...56B} (which allows for an accurate estimation of the threshold mass to prompt collapse in binary NS mergers, using only the knowledge of the stellar parameters of the maximum-mass TOV models of a given EoS) is explained qualitatively using just three ingredients: i) differentially rotating equilibrium models, ii) the existence of an axisymmetric stability limit, and iii) the existence of the linear relation $J_\mathrm{merger}(M_\mathrm{tot})$, which has a different slope than $J(M_\mathrm{stab})$.
These three ingredients uniquely lead to the linear relation $k(C_\mathrm{max})$. 

The qualitative and quantitative agreement between $k^\mathrm{eq}$ and $k^\mathrm{sim}$ is remarkable given the coarse assumptions made for computing $k^{\rm eq}$, i.e. assuming stationarity and axisymmetry, neglecting thermal effects and assuming a simple  rotation law.

\subsubsection{The $M_{\rm stab}(J)$ relation}
For stable equilibrium models that are relevant for our study the relation $J(M_\mathrm{stab})$ shown in Figs. \ref{fig:stablimittm1} and \ref{fig:stab} is almost linear in the range defined by $J=5$ and the intersection with $M_\mathrm{tot}(J_\mathrm{merger})$. In the following, we approximate its inverse $M_\mathrm{stab}(J)$ by the linear relation 
\begin{equation}
M_\mathrm{stab}(J)=c_1 J+c_2,
\label{mstabJ}
\end{equation} and explore the dependence of the parameters $c_1$ and $c_2$ on the EoS.

\begin{figure}
  \includegraphics[width=8.8cm]{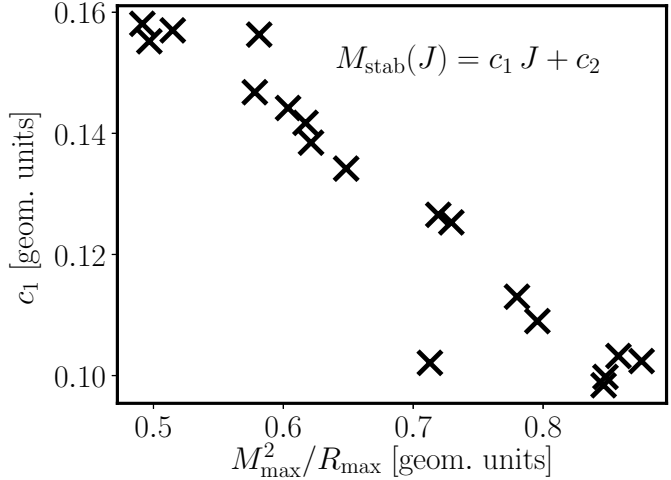}
\caption{\label{fig:afit}Slope parameter $c_1$ of a linear least-squares fit
$c_1\,J+c_2$ to $M_\mathrm{stab}(J)$ as a function of $M_\mathrm{max}^2/R_\mathrm{max}$
for different EoSs. For the fit $M_\mathrm{stab}(J)$ is considered in the
range between $J=5$ and the intersection which determines the threshold mass
(see Fig.~\ref{fig:stab}).}
\end{figure}
\begin{figure}
  \includegraphics[width=8.8cm]{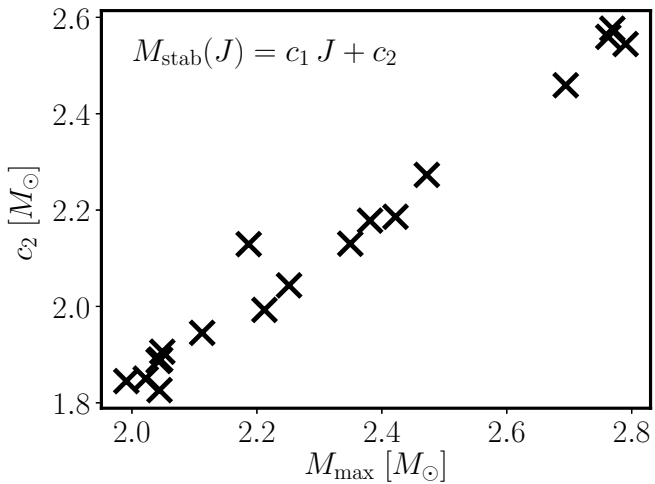}
\caption{\label{fig:bfit}Parameter $c_2$ of a linear least-squares fit $c_1\,J+c_2$ to $M_\mathrm{stab}(J)$ as a function of $M_\mathrm{max}$ for different EoSs. For the fit $M_\mathrm{stab}(J)$ is considered in the range between $J=5$ and the intersection which determines the threshold mass (see Fig.~\ref{fig:stab}).}
\end{figure}

Fig.~\ref{fig:afit} reveals that the slope $c_1$ in (\ref{mstabJ}) has an approximately linear dependence on $M_\mathrm{max}^2/R_\mathrm{max}$, which is a rough measure of the binding energy of the non-rotating maximum-mass configuration. The linear least-squares fit of the data in Fig.~\ref{fig:afit}  is
\begin{equation} 
c_1=-0.165\,(M_\mathrm{max}^2/R_\mathrm{max}) + 0.2412.
\label{c1}
\end{equation}
On the other hand, Fig.~\ref{fig:bfit} reveals that the parameter $c_2$ has an
approximately linear dependence on
$M_\mathrm{max}$. 
The linear least-squares fit of the data in
Fig.~\ref{fig:bfit}
is\begin{equation}
c_2=0.913\,M_\mathrm{max}+0.0133.
\label{c2}
\end{equation}Notice that that practically $c_2 \sim M_\mathrm{max}$, which is consistent with the fact that in the limit of $J=0,$ $M_\mathrm{stab}$ tends to $M_\mathrm{max}$.

The outliers in Fig.~\ref{fig:afit} correspond to the APR EoS at $c_1=0.102$ and to the LS220 EoS at $c_1=0.156$. In Fig.~\ref{fig:bfit} the APR EoS (with $M_\mathrm{max}=2.19~M_\odot$) somewhat deviates from the overall trend. Removing these outliers hardly changes the fits in Eqs.~\eqref{c1} and~\eqref{c2}. In Fig.~\ref{fig:kplot} these two models at $C_\mathrm{max}=0.326$ (APR) and at $C_\mathrm{max}=0.284$ (LS220) do not appear as outliers.

From the above findings, it follows that {\it a given amount of angular momentum
provides more stabilization to less bound NSs}. This behavior will likely  hold
for other rotation laws, but the extent of its validity remains to be explored in more detail. 

Notice that for uniformly rotating NSs we find that $M_\mathrm{stab}^\mathrm{uni}(J)$ is a linear
function of $J$ for $J>2$ (in geometrical units). In the range $2\le J\le4$ the slope of this linear relation scales well with
$M_\mathrm{max}^3/R_\mathrm{max}$ for different EoSs.

\subsubsection{Semi-analytic derivation}
With the relations describing $c_1$ and $c_2$ as functions of $M_\mathrm{max}$ and $R_\mathrm{max}$ (Figs.~\ref{fig:bfit} and~\ref{fig:afit}) we are in the position to arrive at a semi-analytic derivation of the empirical $k=k(C_{\rm max})$ relation, i.e. Eq. (\ref{ksim}). 

We look for an algebraic solution of
the system of equations (\ref{eq:jmerger}) and  (\ref{mstabJ}), i.e. for the intersection of the axisymmetric instability limit $M_\mathrm{stab}(J)$ with the angular momentum available for a given total binary mass, 
 $J_\mathrm{merger}(M_\mathrm{tot})$, which defines the threshold mass 
 $M_\mathrm{thres}$. Eliminating the angular momentum from these two relations, we find
\begin{equation}
M_{\rm thres}=\frac{-c_1 b+c_2}{1-c_1a}.
\end{equation}
Substituting  $c_1$ and $c_2$ from (\ref{c1}) and (\ref{c2}), using $a=4.04$, $b=4.66$ (as derived for the DD2 EoS) and dividing by $M_\mathrm{max}$, we arrive at a relation of the form $k(C_{\rm max},M_{\rm max})$.

\begin{figure}
  \includegraphics[width=8.8cm]{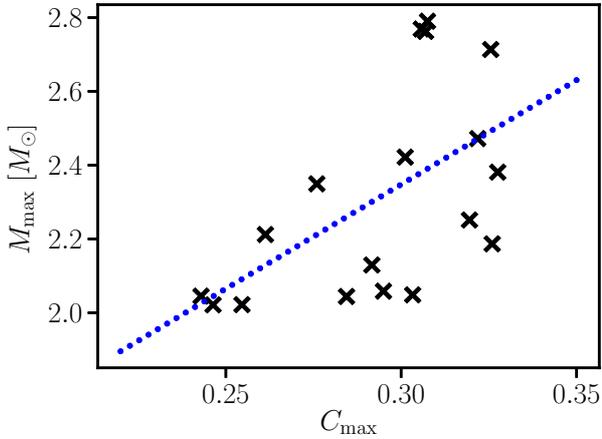}
\caption{\label{fig:mmaxcmax}Maximum mass of non-rotating NSs as function of the compactness of the maximum-mass confirguration of non-rotating NSs. The blue dotted line shows a least-squares fit to the data.}
\end{figure}

In addition, the linear relation $M_\mathrm{max}=5.65~M_\odot\,C_\mathrm{max}+0.65~M_\odot$ roughly describes the relation between $M_\mathrm{max}$ and $C_\mathrm{max}$ for the set of EoSs listed in Tab.~\ref{tab1}, see blue dotted line in Fig.~\ref{fig:mmaxcmax}. Inserting this linear relation in $k(C_{\rm max},M_{\rm max})$, we arrive at a nonlinear expression $k(C_\mathrm{max})$, which is very well approximated by the linear expression
\begin{equation} \label{kan}
k=-3.61 C_\mathrm{max} + 2.44
\end{equation}
in the range $C_\mathrm{max} = \{0.243, 0.328\}$ of the EoS models shown in Tab.~\ref{tab1}
(taking a first-order Taylor expansion around the mid-point $C_\mathrm{max} = 0.2852$).

In the relevant range $0.23<C_{\rm max}<0.34$ Eq.~(\ref{kan}) reproduces the values of $k$ given by the empirical fit~(\ref{ksim}) to the simulation data within 6\%, which is of similar order as the uncertainty in Eq.~(\ref{ksim}). The fit to our equilibrium models, Eq.~(\ref{eq:keq}), is reproduced with a precision of better than one per cent. Alternatively, a relation similar to Eq.~(\ref{kan}) is obtained without using the $M_\mathrm{max}(C_\mathrm{max})$ relation in Fig.~\ref{fig:mmaxcmax} by noticing that $k(C_\mathrm{max}, M_\mathrm{max})$ depends stronger on $C_\mathrm{max}$ than on $M_\mathrm{max}$ and taking a first-order Taylor expansion in terms of $C_\mathrm{max}$ at the mid-point of the range of values for $C_\mathrm{max}$, while fixing $M_\mathrm{max}$ to the mid-point value in the range of values for $M_\mathrm{max}$.

\subsubsection{Radii and the $f_\mathrm{thres}(M_\mathrm{thres})$ relation}
\begin{figure}
  \includegraphics[width=8.8cm]{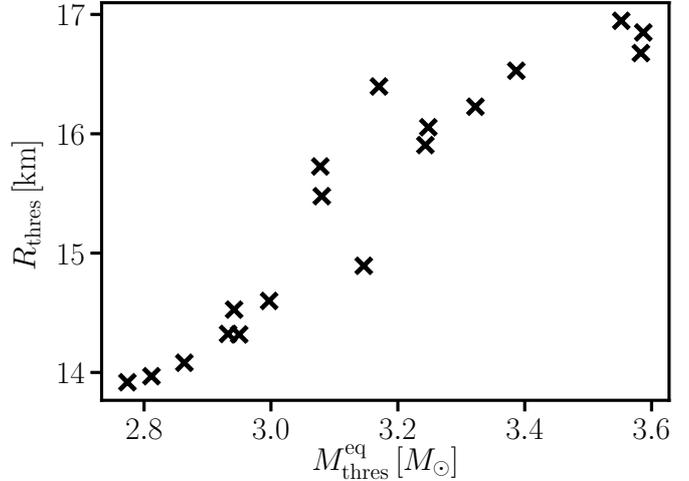}
\caption{\label{fig:radii}Equatorial radii of the threshold stellar equilibrium
model as function of the threshold mass $M_\mathrm{thres}^\mathrm{eq}$ for
different EoSs.}
\end{figure}

The rotating equilibrium models also provide the radii $R_\mathrm{thres}$
of the stars with mass $M_\mathrm{thres}^\mathrm{eq}$ at the threshold to
black hole collapse, and these are listed in Table~\ref{tab1} for each EoS.
Figure~\ref{fig:radii} shows $R_\mathrm{thres}$
 as function of $M_\mathrm{thres}^\mathrm{eq}$ for different EoSs, and one
recognizes a coarse (roughly linear) relation between $R_\mathrm{thres}$
and $M_\mathrm{thres}^{\rm eq}$ (the outliers at $M_\mathrm{thres}^\mathrm{eq}\sim 3.15~M_\odot$ are the ppEoSa, TMA, ppH4 and ppMPA1 EoSs, cf. Tab.~\ref{tab1}). This relation corroborates another empirical
finding of NS merger simulations concerning the gravitational-wave frequency
of the dominant postmerger oscillation.

In \citet{2014PhRvD..90b3002B,Bauswein:2015wsa} we considered the dominant
oscillation frequency of the postmerger remnants as a function of the total
binary mass. The gravitational-wave frequency increases with increasing binary mass
until it reaches a maximum terminal value at the threshold to prompt
collapse  (see Fig.~1 in \citet{2014PhRvD..90b3002B}). The
maximum gravitational-wave frequency $f_\mathrm{thres}$ thus marks the threshold
to direct black-hole formation and it was found to correlate with $M_\mathrm{thres}$, see Fig.~2 and Eq. (1) in \citet{2014PhRvD..90b3002B}. This empirical relation is the key to a procedure
that employs detections of the dominant postmerger gravitational-wave frequency
at lower binary masses to infer the maximum mass of non-rotating NSs and
other stellar properties of the maximum-mass TOV solution \citep{2014PhRvD..90b3002B,Bauswein:2015wsa,2016EPJA...52...56B}.

The existence of the relation between $f_\mathrm{thres}$ and $M_\mathrm{thres}$
found in \citet{2014PhRvD..90b3002B}  can be explained qualitatively by the empirical relation between $R_\mathrm{thres}$ and  $M_\mathrm{thres}$ shown in Fig.~\ref{fig:radii}. For non-rotating stars, the frequency of the fundamental $l=|m|=2$
oscillation mode (the dominant oscillation mode in the postmerger
remnants \citep{2011MNRAS.418..427S}) scales with $\sqrt{\frac{M}{R^3}}$ \citep[see][]{1998MNRAS.299.1059A}. But, also
for rotating merger remnants with a complex velocity profile, the dominant
gravitational-wave frequency scales well with the radius of the remnant for
a fixed binary mass (see Fig.~13 in \citet{2012PhRvD..86f3001B}). It follows that for the model at the threshold to collapse, the empirical relation between $R_\mathrm{thres}$ and $M_\mathrm{thres}$
simplifies the dependence of the dominant postmerger frequency to a relation between $f_\mathrm{thres}$
and $M_\mathrm{thres}$ only. We plan to revisit the above qualitative arguments in the future, using perturbative calculations 
of the frequency of the fundamental $l=|m|=2$ modes in  equilibrium models that resemble merger remnants.

\section{Conclusions}
We construct equilibrium models of differentially NSs for a large sample of EoSs. For these configurations we identify the maximum mass that can be supported against gravitational collapse for a given amount of angular momentum. For a given EoS this defines a stability limit. We interpret these results in the context of NS mergers by assuming that merger remnants are qualitatively represented by our equilibrium models. Comparing the masses and angular momenta at the stability limit of equilibrium models to corresponding parameters of binary mergers, we determine a threshold mass which marks the onset of instability of equilibrium merger models. Within our model this threshold mass corresponds to the threshold binary mass for prompt collapse of the merger remnant.

Investigating a large sample of EoSs we find that the threshold mass of equilibrium models follows qualitatively the same behavior as the binary threshold mass for prompt collapse found in merger simulations. Specifically, the threshold mass can be well described as a fraction $k$ of the maximum mass of non-rotating NSs with $k$ being a linear function of the maximum compactness $C_{\rm max}$  of non-rotating NSs. This is an important finding because it confirms in a broader context and for a larger sample of EoSs the collapse behavior of merger remnants.  By this, it emphasizes the possibility of inferring the maximum mass of non-rotating NSs from future gravitational-wave detections \citep{2013PhRvL.111m1101B,2016EPJA...52...56B}. In turn, employing available constraints on $M_\mathrm{max}$ and $R_\mathrm{max}$ the relation for $M_\mathrm{thres}=k(C_\mathrm{max})\,M_\mathrm{max}$ will be useful to interpret future simultaneous detections of electromagnetic radiation (from radioactively powered transients or short GRBS) and gravitational waves providing the binary masses. In particular, it will determine the nature of the underlying merger remnant (black hole or NS remnant) leading to the electromagnetic signal.

In constructing the equilibrium models we made a number of simplifying assumptions: the models are assumed to be stationary and axisymmetric, thermal effects are neglected and  a simple rotation law is adopted. Still, we were able to reproduce semi-analytically the empirical relation $k(C_{\rm max})$  found previously through merger simulations \citep{2013PhRvL.111m1101B,2016EPJA...52...56B}. This demonstrates that this relation is indeed very robust and  has only a weak dependence  on thermal effects, on  deviations from axisymmetry and stationarity and on the details of the differential rotation law. The threshold mass for collapse during NS mergers is mostly depending on the compactness of
the maximum-mass non-rotating model (but it could depend on additional factors that are not included in the present study, such as the intrinsic spin or the magnetic field if such effects are strong).

The existence of the  $k(C_{\rm max})$ relation can be traced back to the existence
of the roughly linear, EoS-insensitive relation $J_\mathrm{merger}(M_\mathrm{tot})$ for the angular momentum in a merger remnant for given total binary mass.  This relation has a significantly different slope with respect to the angular momentum of marginally stable models for a given EoS, $J(M_\mathrm{stab})$ (which for the EoSs we examined is also a roughly linear relation in the range of angular momenta that are typical for NS merger remnants). The intersection between the two relations leads directly to  $k(C_\mathrm{max})$.
 
In reproducing  $k(C_\mathrm{max})$ semi-analytically, we also found that at high angular momentum  $M_\mathrm{stab}(J)$ is a linear  function with slope  $\sim M_\mathrm{max}^2/R_\mathrm{max}$  which implies that  a given amount of angular momentum provides more stabilization to less bound NSs. Furthermore, we found a roughly linear relation between the radius and mass of the models at the threshold to collapse for various EoSs. In conjunction with the dependence of the fundamental quadrupole oscillation mode of NSs on the average density, this could provide a qualitative understanding of the empirical relation $f_\mathrm{thres}(M_\mathrm{thres})$ between the dominant postmerger gravitational wave frequency and the threshold mass for collapse, which was discovered in \citet{2014PhRvD..90b3002B}. This latter relation is useful for estimating the stellar properties of the maximum-mass configuration of non-rotating NSs using postmerger gravitational-wave observations of at least two lower-mass models, following an extrapolation procedure introduced in detail in  \citet{2014PhRvD..90b3002B}.

In the future we plan to investigate in more detail the $f_\mathrm{thres}(M_\mathrm{thres})$ relation using perturbative results for the frequency of the fundamental quadrupole oscillation of differentially rotating NSs. Ultimately, we aim at a qualitative or possibly even quantitative understanding of the empirical relation between the dominant postmerger gravitational-wave frequency and the radius of non-rotating NSs, as found in \citet{2012PhRvL.108a1101B,2012PhRvD..86f3001B}. 

\section*{Acknowledgements}
We thank Matthias Hempel for providing EoS tables. We acknowledge support by the Klaus Tschira Foundation and the Deutsche Akademische Austauschdienst through the IKY-DAAD IKYDA 2016 grant (DAAD No. 57260023). Computing time was provided in part by the GWAVES\#pr002022 allocation on the ARIS computer of GRNET in Athens.

\bibliographystyle{mnras}
\bibliography{references} 




%
%


\bsp	
\label{lastpage}
\end{document}